\documentclass[fleqn,usenatbib]{mnras}

\usepackage{newtxtext,newtxmath}
\usepackage{adjustbox}
\usepackage{multirow}
\usepackage{booktabs}
\usepackage[super]{nth}
\usepackage{subfig}
\usepackage[T1]{fontenc}
\usepackage{xcolor}
\usepackage{soul}
\DeclareRobustCommand{\VAN}[3]{#2}
\let\VANthebibliography\thebibliography
\def\thebibliography{\DeclareRobustCommand{\VAN}[3]{##3}\VANthebibliography}

\usepackage{graphicx}	
\usepackage{amsmath}	

\title[Blazar Classification]{Gamma-ray Blazar Classification using Machine Learning with Advanced Weight Initialization and Self-Supervised Learning Techniques}

\author[Bhatta et al.]{
Gopal Bhatta,$^{1}$ \thanks{g.bhatta@ia.uz.zgora.pl}
Sarvesh Gharat,$^{2}$ \thanks{sarveshgharat19@gmail.com}
Abhimanyu Borthakur$^{3}$
and Aman Kumar$^{4}$
\\
$^{1}$Janusz Gil Institute of Astronomy, University of Zielona Góra, ul. Szafrana 2, 65-516 Zielona Góra, Poland\\
$^{2}$Centre for Machine Intelligence and Data Science, Indian Institute of Technology Bombay, 400076, Mumbai, India\\
$^{3}$Department of Electronics and Communication Engineering, Manipal Institute of Technology, 576104, Karnataka, India\\
$^{4}$Tezpur University, Tezpur, Assam 784028, India.
}

\date{Accepted XXX. Received YYY; in original form ZZZ}

\pubyear{2023}

\begin{document}
\label{firstpage}
\pagerange{\pageref{firstpage}--\pageref{lastpage}}
\maketitle

\begin{abstract}
Machine learning has emerged as a powerful tool in the field of gamma-ray astrophysics. The algorithms can distinguish between different source types, such as blazars and pulsars, and help uncover new insights into the high-energy universe. The Large Area Telescope on-board the Fermi Gamma-ray telescope has significantly advanced our understanding of the Universe. The instrument has detected a large number of gamma-ray emitting sources, among which a significant number of objects have been identified as active galactic nuclei. The sample is primarily composed of blazars; however, more than one-third of these sources are either of an unknown class or lack a definite association with a low-energy counterpart. In this work, we employ multiple machine learning algorithms to classify the sources based on their other physical properties. In particular, we utilized smart initialisation techniques and self-supervised learning for classifying blazars into BL Lacertae objects (BL Lac, also BLL) and flat spectrum radio quasars (FSRQ). The core advantage of the algorithm is its simplicity, usage of minimum number of features and easy deployment due to lesser number of parameters without compromising on the performance along with increase in inference speed (at least $7$ times more than existing algorithms). As a result, the best performing model is deployed on multiple platforms so that any user irrespective of their coding background can use the tool. The model predicts that out of the $1115$ sources of uncertain type in the 4FGL-DR3 catalog, $820$ can be classified as BL Lacs, and $295$ can be classified as FSRQs.
\end{abstract}





\section{Introduction}
Blazars, belonging to the class of active galactic nuclei (AGNs), stand out as some of the most luminous and exceptionally variable sources in the Universe. 
These sources are recognized for their high luminosity, broad-spectrum emissions, and significant rapid variability across a wide range of the electromagnetic spectrum \citep[see e.g.,][]{2021ApJ...923....7B,2020ApJ...891..120B,2018A&A...619A..93B}. These exceptional characteristics are frequently associated with the emission boosted by Doppler effects from the relativistic outflows originating near the central engine \citep{1995PASP..107..803U,2017ApJ...846...98J}.
Conventionally, these objects are typically divided into two main groups: BL Lacs and FSRQ.

The primary distinction between these two categories lies in the fact that BL Lacs typically display either no or very faint emission line spectra, whereas FSRQs commonly exhibit broad emission lines and and their synchrotron peak is at lower frequencies. While FSRQs are more powerful sources, BL Lacs belong to an extreme class characterized by an excess of high-energy emissions, ranging from hard X-rays to TeV energies. In leptonic models of blazar, this excess arises from synchrotron and inverse-Compton (IC) processes. Their seemingly low luminosity may be attributed to the absence of strong circumnuclear photon fields and relatively low accretion rates \cite{2011MNRAS.414.2674G}
\begin{figure*}
    \centering
    \includegraphics[scale = 0.60]{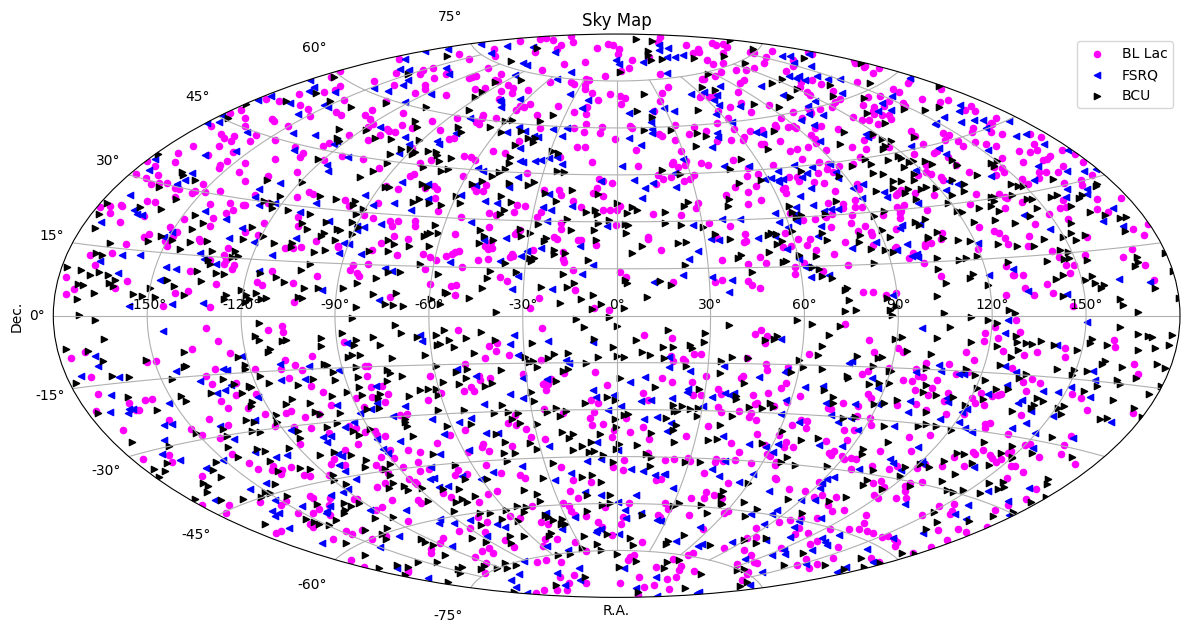}
    \caption{Sky map showing population of BL Lacs, FSRQs and BCUs from Fermi/LAT recent catalog of AGN  4FGL LAC-DR3}
    \label{fig:Sky_Map}
\end{figure*}

Blazars are known for emitting non-thermal radiation across a wide spectrum, spanning from radio waves to TeV gamma-rays. The spectral energy distributions (SEDs) of blazars exhibit two distinct peaks. The first peak, observed in the infrared to soft X-ray energy range, originates from synchrotron emission. In contrast, the second peak, situated in the hard X-ray to gamma-ray region, is associated with inverse Compton (IC) radiation, as per the leptonic model \citep[IC; see][for a recent review]{2019Galax...7...20B}. The photons involved in IC scattering can either arise from the same group of electrons responsible for generating the synchrotron peak, as explained by the synchrotron self-Compton (SSC) model \citep{1992ApJ...397L...5M,1996ApJ...461..657B}, or they can originate from external sources, including, but not limited to, the accretion disk \citep{dermer1993model}, the broad line region \citep{1994ApJ...421..153S}, or the dust torus \citep{2000ApJ...545..107B}

An alternative approach to categorizing blazars involves examining their Spectral Energy Distributions (SEDs) and considering the synchrotron peak frequency ($\nu_s$). Following this approach, blazars can be classified into three categories: high-  (HSP; $\nu_s > 10^{15}$ Hz), intermediate- (ISP; $10^{14} < \nu_s < 10^{15}$ Hz), or low- (LSP; $\nu_s < 10^{14}$ Hz) synchrotron peaked blazars. It is noteworthy that, in this classification approach, FSRQs primarily fall under the category of LSP. In the unifying scheme known as the blazar sequence, the bolometric luminosity decreases as the sources transition from FSRQs to HSP sources, whereas gamma-ray emissions increase \citep{2017MNRAS.469..255G,1998MNRAS.299..433F}. 

The Fermi Large Area Telescope (LAT), in operation since 2008, conducts a continuous survey of the entire sky, identifying gamma-ray sources within the energy range spanning from tens of MeV to the TeV range \cite{2009ApJ...697.1071A}.
Among the extragalactic $\gamma$-ray sources observed by the Fermi $\gamma$-ray space telescope, blazars constitute the largest population \cite{2020ApJS..247...33A}.
A significant number, roughly one-third, of the blazar candidates identified by Fermi-LAT up to this point belong to uncertain category.
Blazar Candidates of Uncertain type (BCU) are blazar candidates that do not clearly fit into one of the established blazar subtypes, such as BL Lac or FSRQ. 
In other words, these blazar candidates may have characteristics that make it difficult to definitively classify them into a specific blazar subtype. One often needs additional data and analysis to determine the precise nature of these objects. In some cases, this might prove to be an expensive and challenging task, especially when dealing with sources that could potentially be of the BL Lac type, which inherently have weak or no emission lines.

As an alternative approach, several authors have recently turned to various ML algorithms to classify BCUs into BL Lacs and FSRQs. For example, \cite{2023MNRAS.525.1731C} used the MICE and k-nearest neighbors (kNN) algorithms to initially fill in missing variables, such as redshift and the highest energy, and subsequently classified AGNs into either BL Lacs or flat spectrum radio quasars (FSRQs) using multiple algorithms based on the SuperLearner. In another study, \cite{2023MNRAS.519.3000S} employed Artificial Neural Networks, XGBOOST, and LIGHTGBM algorithms to classify BCUs into 825 BL Lac candidates and 405 FSRQ candidates, along with 190 sources that remained without a clear prediction. \cite{2023ApJ...946..109A} used multiple supervised ML algorithms and classified a sample of 1,115 BCUs into 610 BL Lac objects and 333 FSRQs. Additionally, \cite{2022JCAP...04..023B} used Bayesian neural networks for the classification of Fermi-LAT blazar candidates into BL Lacs and FSRQs while also estimating associated uncertainties. In a separate study, \cite{2020MNRAS.493.1926K} employed a supervised machine learning method based on an artificial neural network on a sample of 1,329 BCUs, predicting that 801 sources are BL Lacs, 406 are FSRQs, and 122 remain unclassified. Finally, \cite{2019ApJ...887..134K} employed supervised machine-learning algorithms to generate predictive models, classifying BCUs into 724 BL Lac-type candidates, 332 FSRQ-type candidates, and 256 without a clear prediction.

In this study, we use data from the 4th Fermi catalog (4FGL, \cite{2020ApJS..247...33A}) and employ machine learning-based classification methods to distinguish between BCUs, classifying them as either BLLs or FSRQs. The method majorly depends on using of smarter ways of initialising weights and employing self-supervised learning. It has been observed that from all the employed techniques, bias initialisation with soft voting happens to perform the best while giving the accuracy of $93\%$ and macro average F1 score of $0.914$. Other methods deliver a similar performance with accuracy ranging from $88.6\%-91.5\%$. The core advantage of our method over other available methods is its simplicity making it easy to deploy and speed up the inference speed, reproduciblity that helps in reproducing the results and also adopt the method directly in case of predictions in bulk, and finally using minimum features resulting in a method that uses all possible sources from the catalog without compromise. 

In this paper, we organize our content as follows: Section 2 introduces the dataset and outlines our classification methods, including both vanilla architectures and intelligent modifications, as well as our training and testing strategies. Section 3 presents our results and discusses their implications. Finally, in Section 4 we outline our conclusions.

\section{Methodology}
\subsection{Data Collection and Processing}
In this study, we make use of Fermi's fourth catalog of active galactic nuclei (AGNs) data release 3 (4LAC-DR3 \footnote{\url{https://fermi.gsfc.nasa.gov/ssc/data/access/lat/4LACDR3/}}; \citealt{2022ApJS..263...24A}), which is based on data accumulated over 12 years and contains over $6600$ sources. The catalog comprises $1458$ BL-Lac objects, $792$ FSRQs, and $1493$ BCUs, which are shown in the sky map in Figure \ref{fig:Sky_Map} by the symbols in colors magenta, blue and black, respectively. 
The catalog consists of 3407 sources, each defined by a total of 41 observational features. However, 35 features are provided without any missing values. The feature 'SED\_class' has the highest number of missing values (989), while the 'Counterpart\_Catalog' feature has the least, with only 20 missing values.
However, for this study we only make use of clean samples (refer to \cite{2020ApJS..247...33A} for more information), hence reducing the sample size to $1335$ BL-Lac objects, $670$ FSRQs, and $1115$ BCUs respectively. Further, we observe that although considering all the features can result in better results, it minimizes the number of samples for which we can make predictions due to the missing values, and imputation of it can result in artificial bias making things erroneous. Hence, along with referring to important contributing features in \cite{2023ApJ...946..109A} and the availability of data for different features, we proceed with $7$ features namely "PL\_Index, nu\_syn, LP\_Index, Pivot\_Energy, Frac\_Variability, Variability\_Index, and nuFnu\_syn" \footnote{A description of the observed features is available in Table A1 of \cite{2022ApJS..263...24A}, providing detailed information on the characteristics under consideration.}.
Next, the collected data is divided into train, test, and validation sets in the ratio of $80:10:10$. To have a fair evaluation of the proposed methods, we ensure that the test data consists of data that has never been seen by the algorithm previously. Further, all features except PL\_Index, LP\_Index, and Frac\_Variability underwent log transformations, which ensures that the large values of parameters do not explicitly distort the learning of the model.

\begin{figure*}
    \centering
    \subfloat[Bias Initialisation with Soft Voting]{\includegraphics[scale = 0.55]{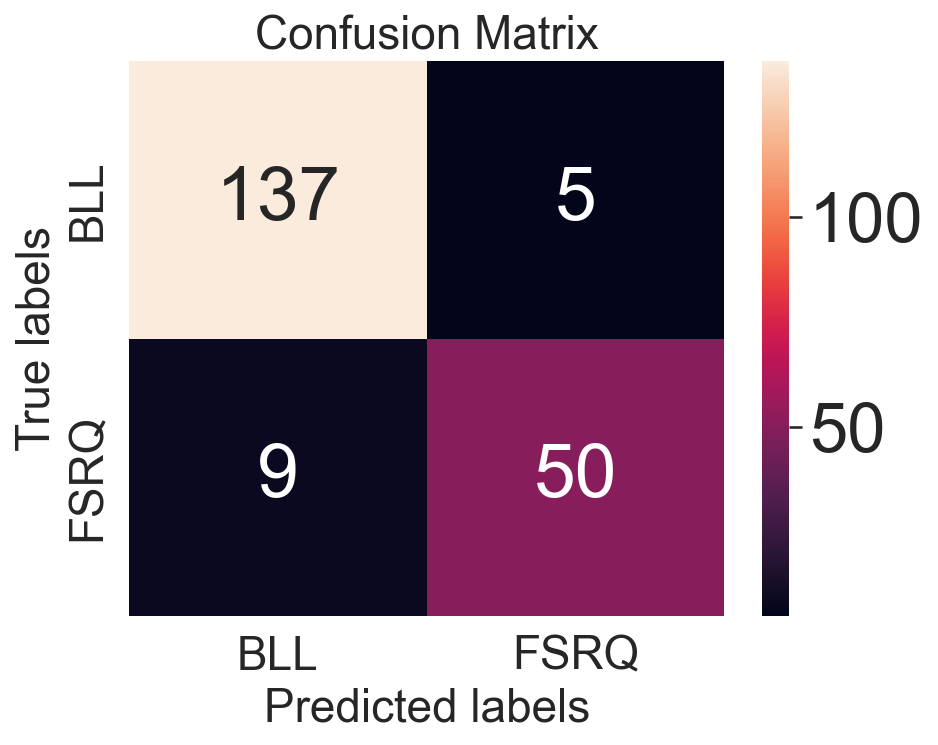}\label{fig:conf_model1}} \quad
    \subfloat[Supervised Greedy Pretraining]{\includegraphics[scale = 0.55]{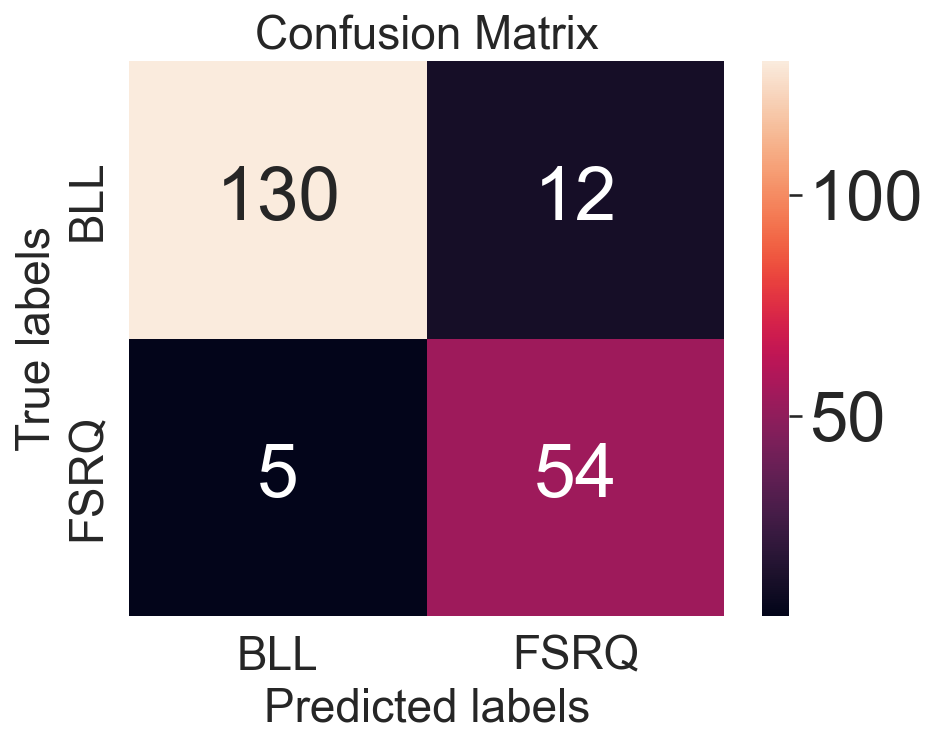}\label{fig:conf_model2}}\quad
    \subfloat[Unsupervised Greedy Pretraining]{\includegraphics[scale = 0.55]{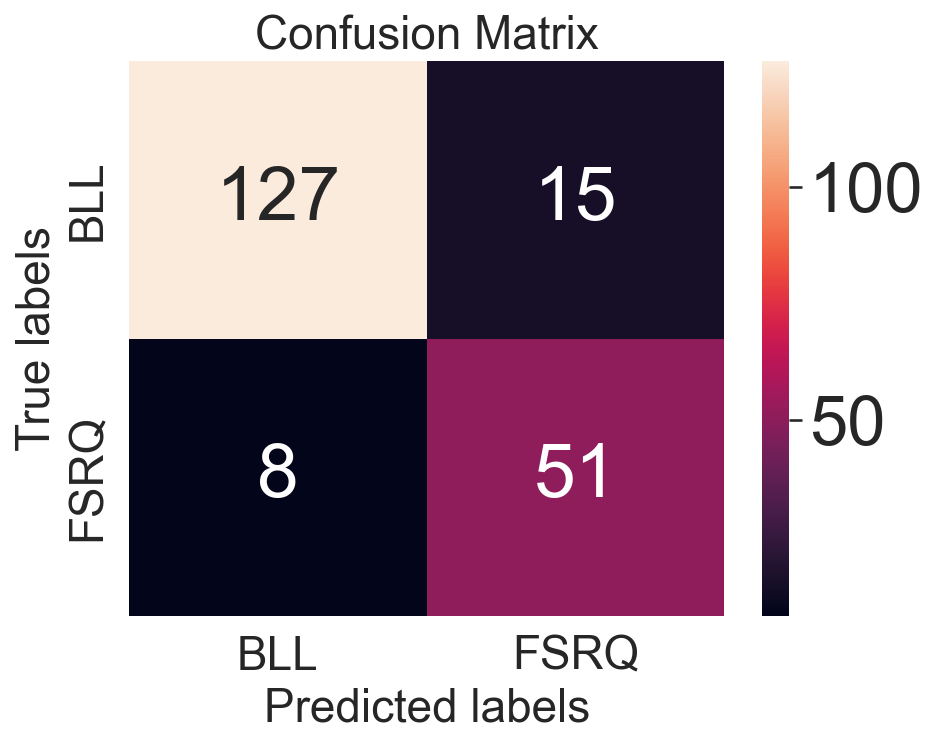}\label{fig:conf_model3}}\\ 
    \subfloat[SSL Autoencoder Pretext]{\includegraphics[scale = 0.55]{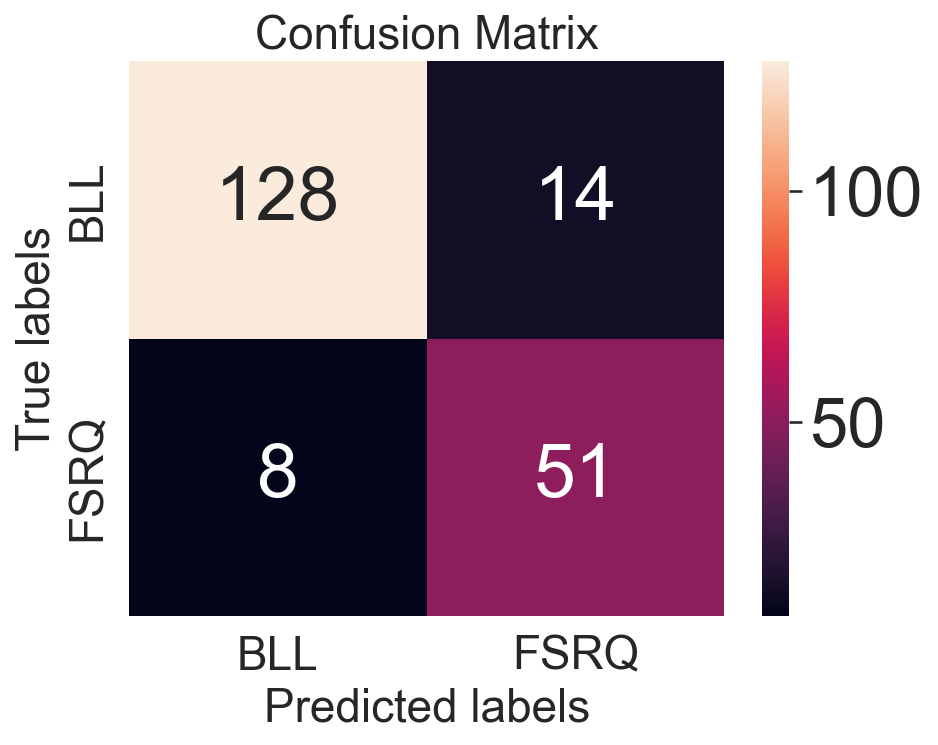}\label{fig:conf_model4}} \quad
    \subfloat[SSL Contrastive Classification]{\includegraphics[scale = 0.55]{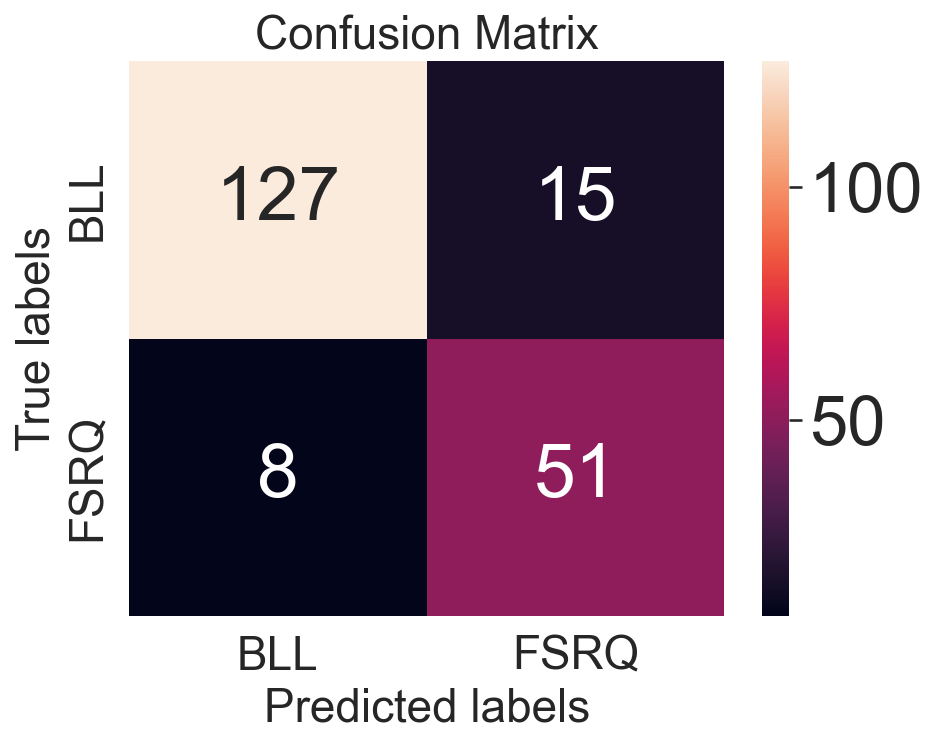}\label{fig:conf_model5}}
    \caption{Confusion Matrix on Testing Data for all the models with X axes denoting the predicted label and Y axes denoting the true label}
    \label{fig:confusion_matrix}
\end{figure*}

\begin{figure*}
    \centering
    \subfloat[Bias Initialisation with Soft Voting]{\includegraphics[scale = 0.55]{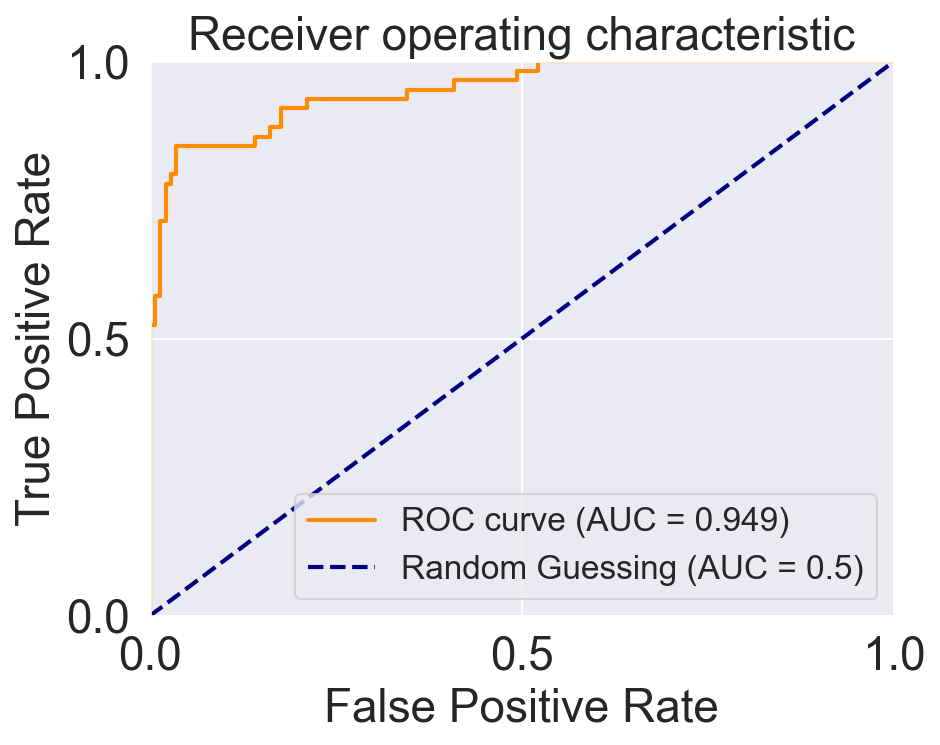}\label{fig:roc_model1}} \quad
    \subfloat[Supervised Greedy Pretraining]{\includegraphics[scale = 0.55]{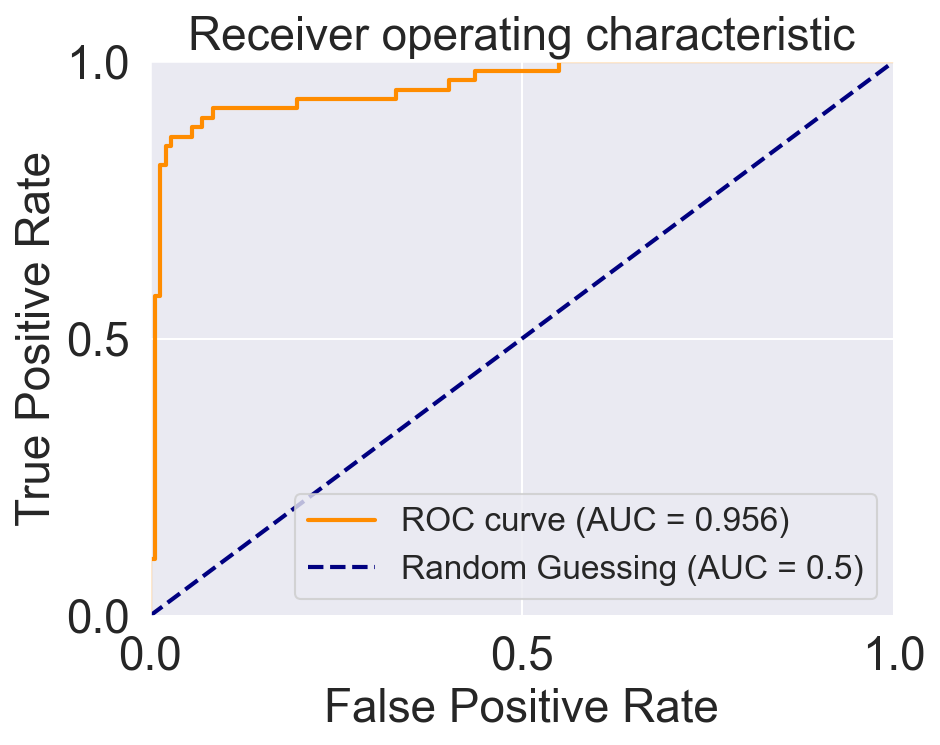}\label{fig:roc_model2}}\\
    \subfloat[Unsupervised Greedy Pretraining]{\includegraphics[scale = 0.55]{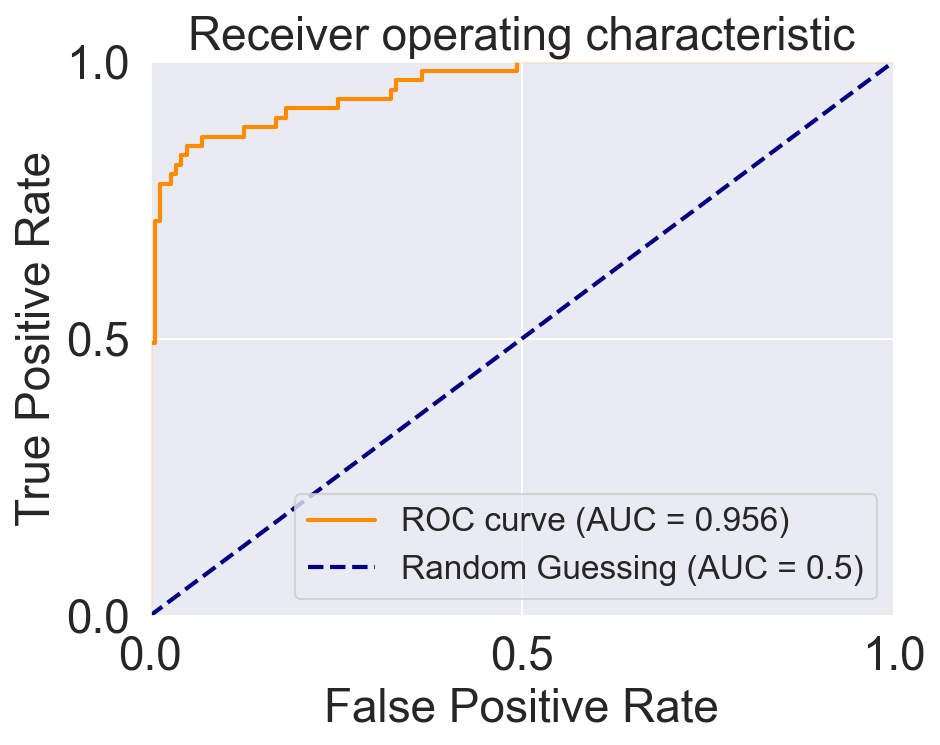}\label{fig:roc_model3}}\quad 
    \subfloat[SSL Autoencoder Pretext]{\includegraphics[scale = 0.55]{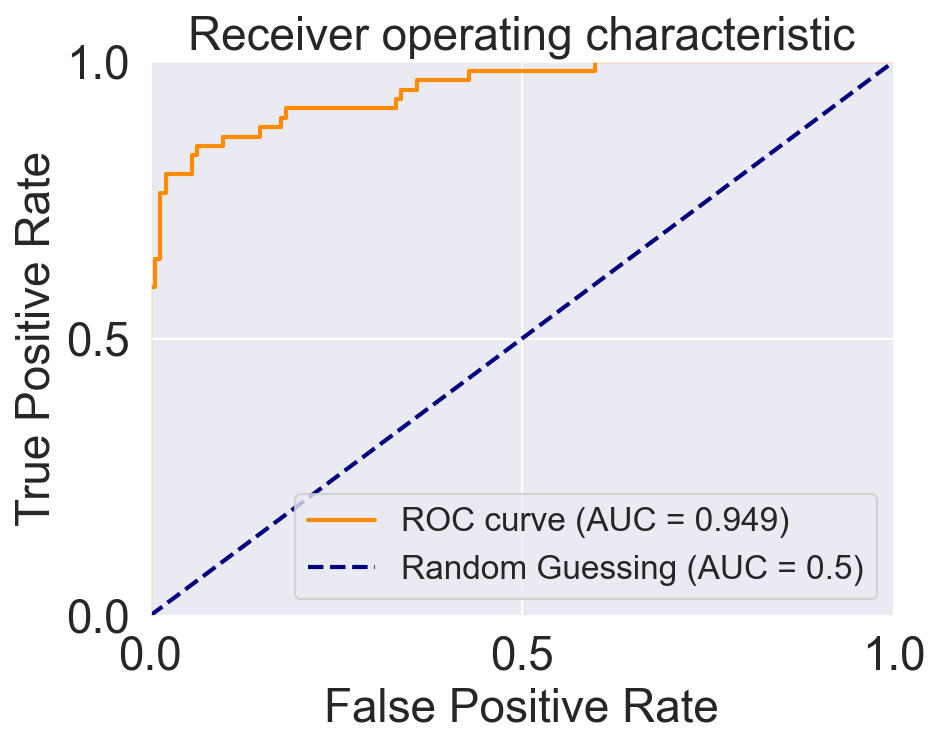}\label{fig:roc_model4}} \quad
    \subfloat[SSL Contrastive Classification]{\includegraphics[scale = 0.55]{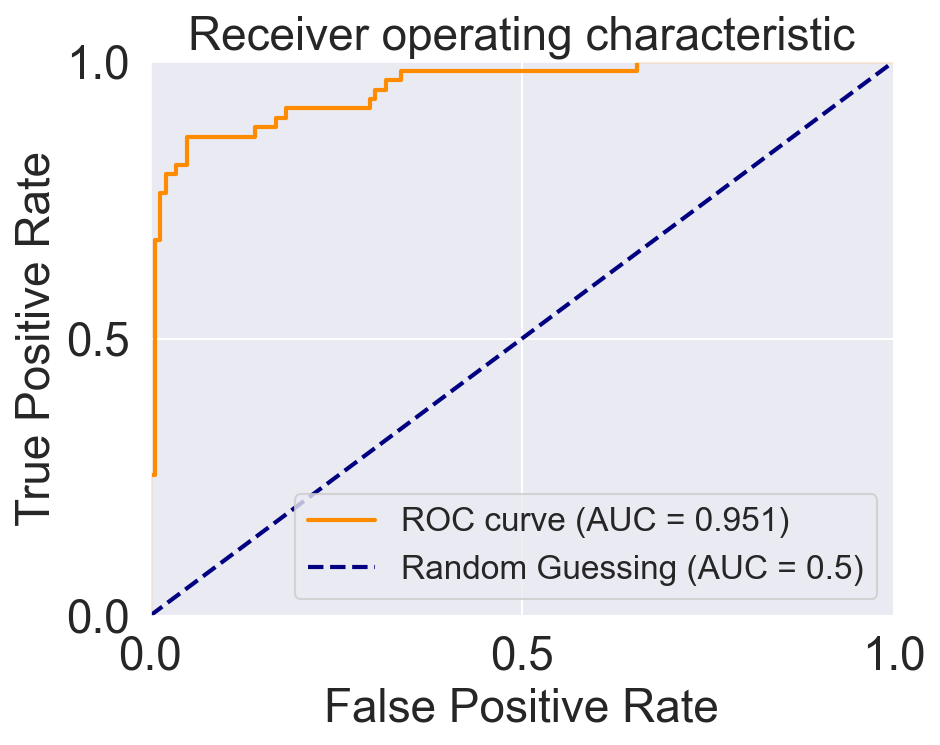}\label{fig:roc_model5}}
    \caption{ROC curves along with corresponding AUC values. The ROC curve is plotted between the False Positive Rate and True Positive rate}
    \label{fig:ROC}
\end{figure*}

\subsection{Model Architecture}
The study boardly explores two different approaches: traditional Artificial Neural Networks (ANNs), and self-supervised learning. The former is divided into three parts, primarily focused on various weight initialization strategies and is elaborated as follows. 
The first and the best-performing architecture uses a bias-initialization technique to initialize the weights in the final layer, a common technique for weight initialization when dealing with imbalanced data-sets. Since the proposed architecture comprises only one hidden layer with $42$ neurons, it has a significantly lower number of parameters, making it computationally efficient. In addition to the initialization, we also introduce an ensemble-based soft voting approach for this architecture - After extensive experimentation and saving the training weights at each epoch, we retrieve the architecture's weights from the \nth{40} and \nth{41} epochs to create two distinct pseudo models. We then evaluate both models on the test data, assigning a weight of $0.1$ to the predictions from the model corresponding to the \nth{40} epoch and a weight of $0.9$ to the predictions from the \nth{41} epoch. The final prediction is calculated as the weighted sum of these predictions, rounded to the nearest whole number.These weight values are hyperparameters tuned through a trial-and-error method.

The remaining two initialisation techniques are applied to an ANN with $2$ hidden layers containing $64$ and $32$ neurons respectively, with a dropout \cite{srivastava2014dropout} of $0.5$ in between them. Making use of dropout in the study, ensures that at any given point while training a particular neuron will remain inactive with probability of $0.5$, hence, ensuring that there isn't any overdependency on a particular neuron thereby encouraging the generalizability of the model.
The second initialization technique is a greedy-based pretraining approach in a supervised fashion.
Here, initially we train the input layer without considering the hidden layer. This provides us with an estimate of the weights that would be best if there were no hidden layer. Next, while keeping the input layer weights constant, we train the algorithm again to estimate the weights of the neurons in the hidden layer.

The third and final approach for initialization is again a greedy-based approach, however in this case we deal in an unsupervised manner \citep{erhan2010does}. This is a widely used technique when dealing with data having a large number of unlabelled data points.
While its purpose is to learn the data distribution for weight initialization, there is no guarantee that this method will achieve optimal performance, even when compared to its vanilla version.
We observe a similar behavior when using an autoencoder-based method to pretrain the network in an unsupervised fashion.

In summary, the first bias-based initialization technique is implemented on a traditional ANN having $1$ hidden layer with the model having $42$ neurons with a dropout of $0.5$ and the other two greedy based approaches are applied to an ANN with $2$ hidden layers containing $64$ and $32$ neurons, respectively,  with a dropout of $0.5$ in between them. As this is a binary classification task, the loss function used for this model is "binary crossentropy" with the activation function being "sigmoid" in the output layer. Using of sigmoid as an activation function restricts the output between $0$ and $1$ allowing us to use a single neuron in the output layer. In our specific case, we represent "BL Lac" by "0" and "FSRQs" by "1". Hence, the more the output is towards one, the more confident is the algorithm in predicting the corresponding input target as FSRQ, similar is the case with BL Lac wherein the output has to be very close to zero.

\begin{figure*}
    \centering
    \subfloat[Bias Initialisation with Soft Voting]{\includegraphics[scale = 0.55]{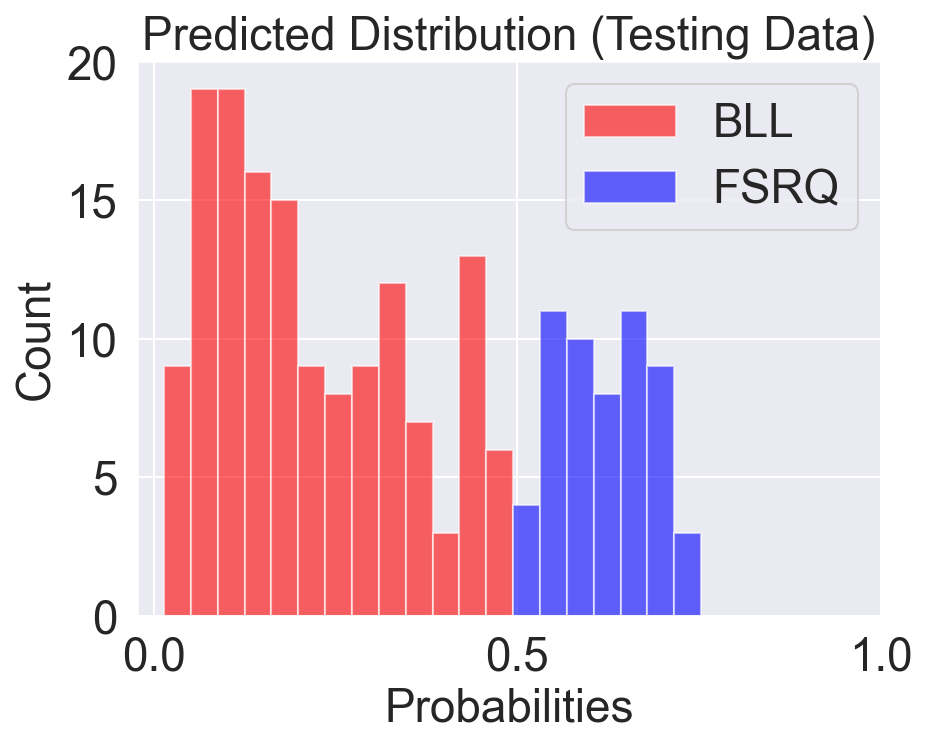}\label{fig:predicted_test_model1}} \quad
    \subfloat[Supervised Greedy Pretraining]{\includegraphics[scale = 0.55]{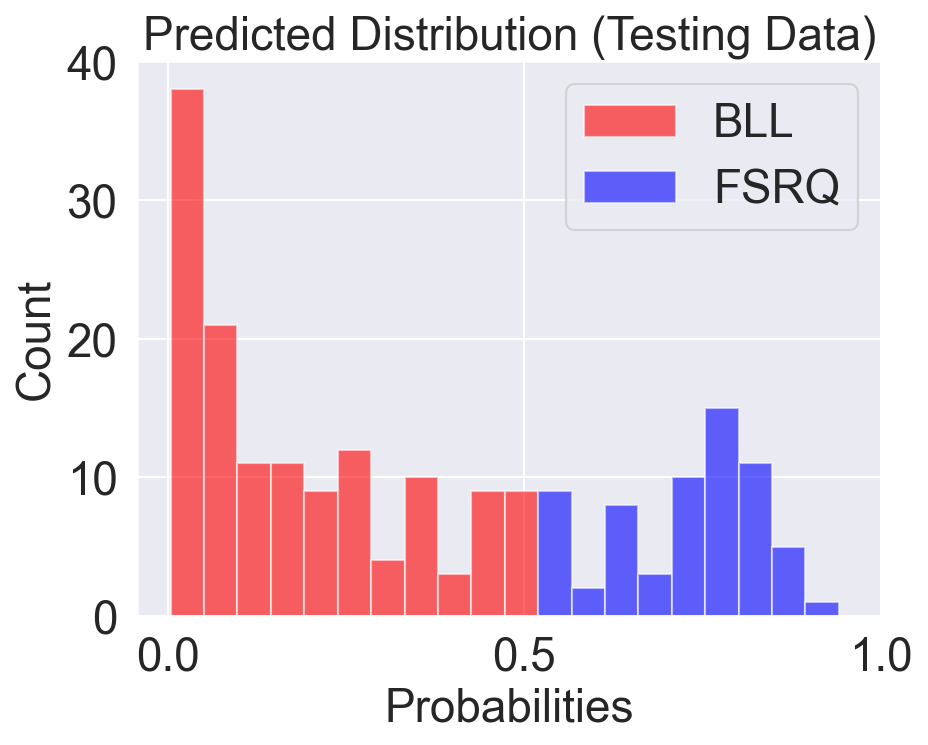}\label{fig:predicted_test_model2}}\quad
    \subfloat[Unsupervised Greedy Pretraining]{\includegraphics[scale = 0.55]{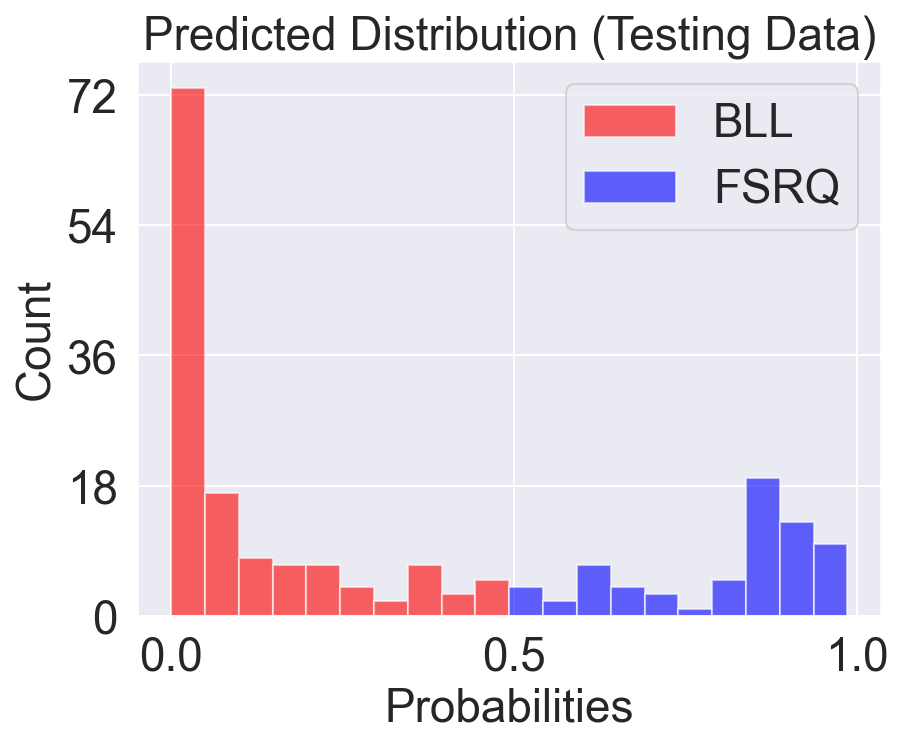}\label{fig:predicted_test_model3}}\\ 
    \subfloat[SSL Autoencoder Pretext]{\includegraphics[scale = 0.55]{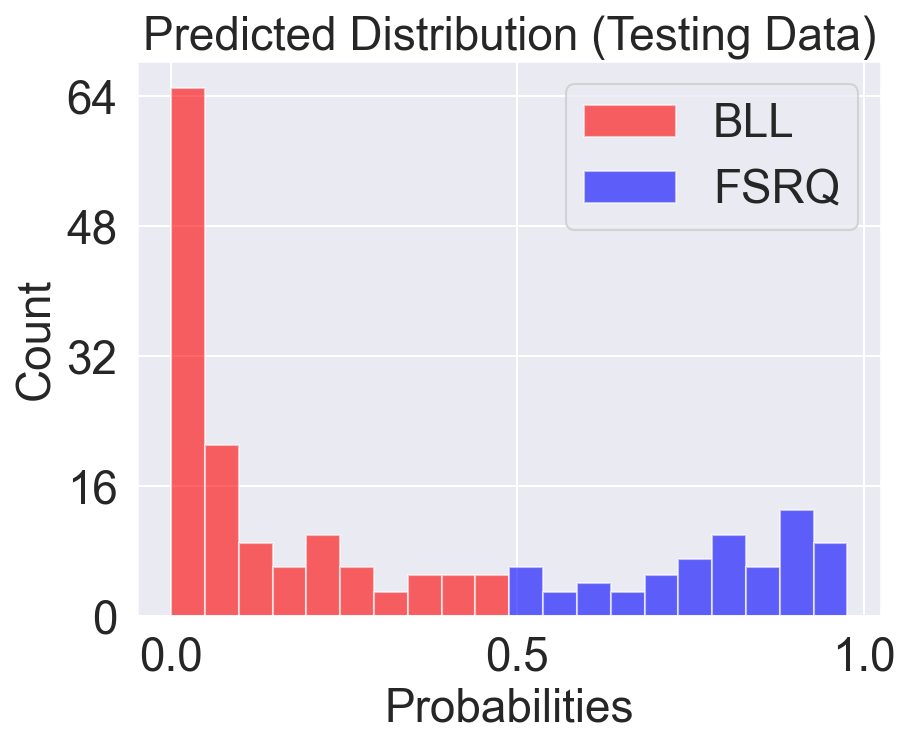}\label{fig:predicted_test_model4}} \quad
    \subfloat[SSL Contrastive Classification]{\includegraphics[scale = 0.55]{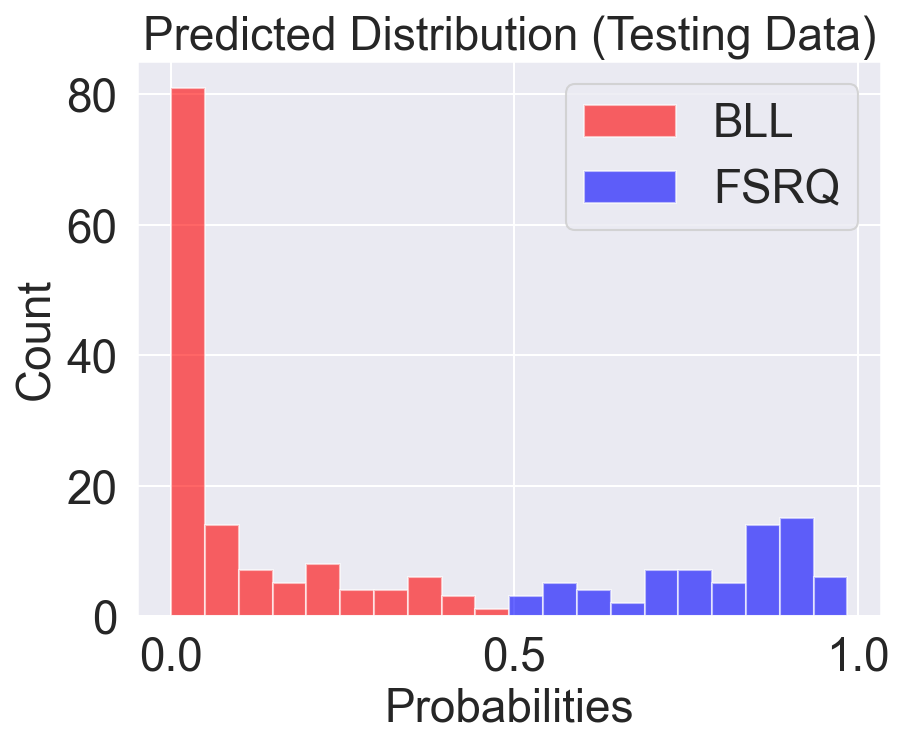}\label{fig:predicted_test_model5}}
    \caption{Distribution of Predicted output on Test Data. Here "0" denotes the BL Lac and "1" denotes FSRQ. The closer the value is to these extremes, the more confident the model is in its prediction.}
    \label{fig:Predicted_Test}
\end{figure*}

\begin{figure*}
    \centering
    \subfloat[Bias Initialisation with Soft Voting]{\includegraphics[ scale = 0.55]{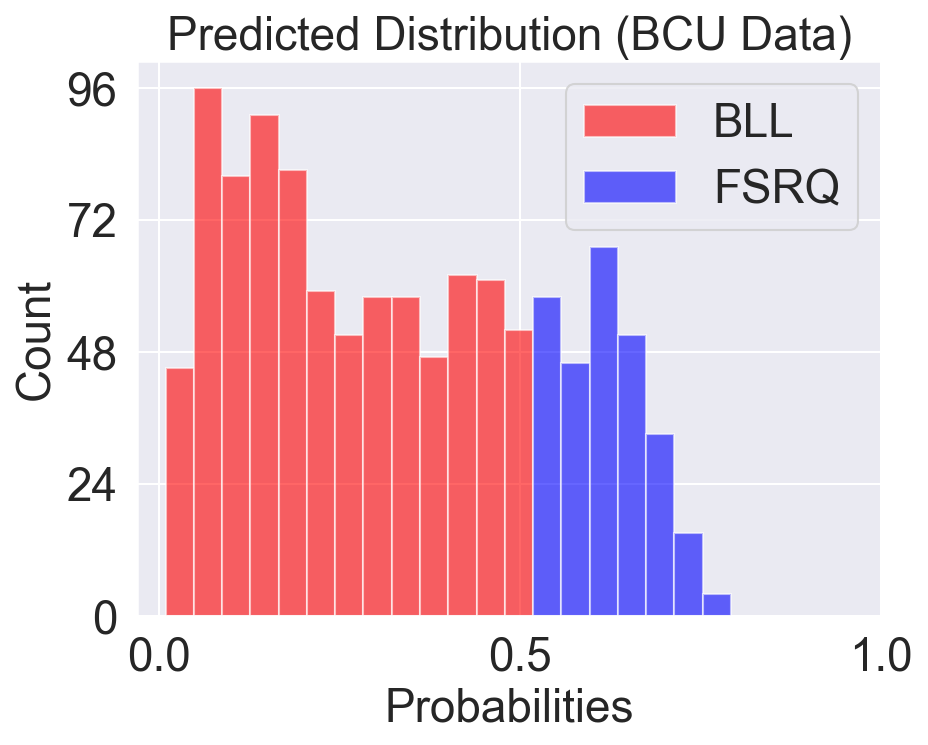}\label{fig:predicted_data_model1}} \quad
    \subfloat[Supervised Greedy Pretraining]{\includegraphics[scale = 0.55]{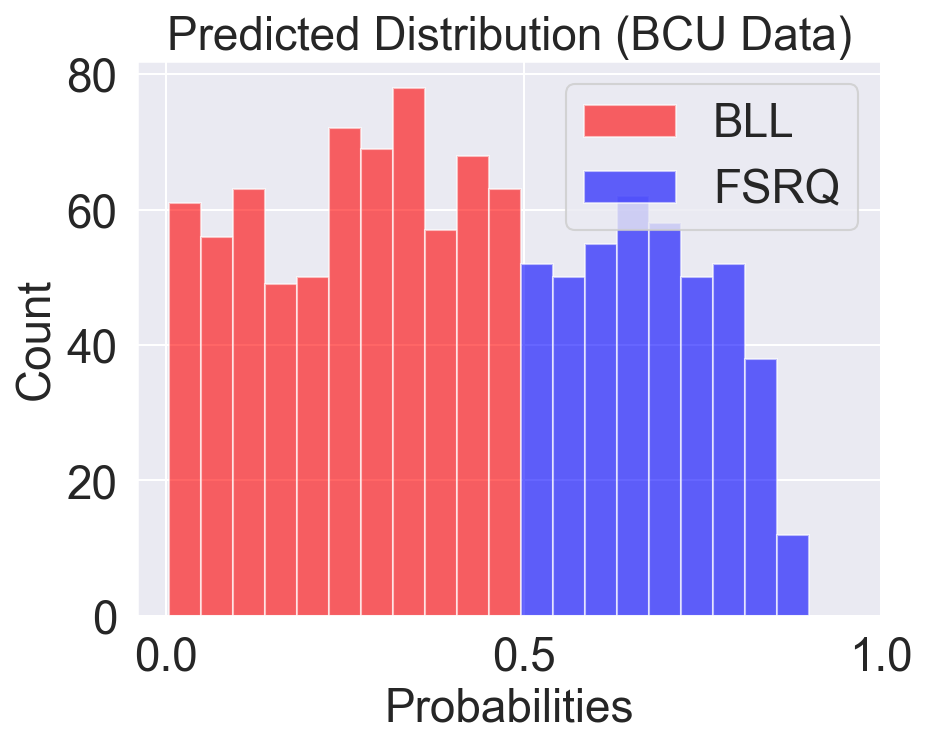}\label{fig:predicted_data_model2}}\\
    \subfloat[Unsupervised Greedy Pretraining]{\includegraphics[scale = 0.55]{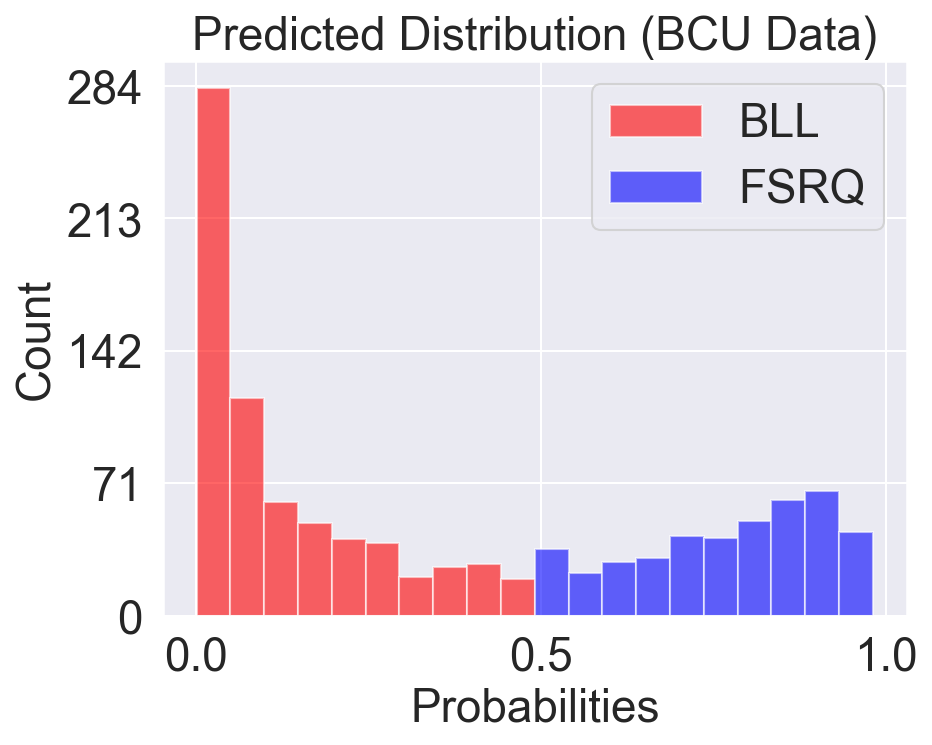}\label{fig:predicted_data_model3}}\quad 
    \subfloat[SSL Autoencoder Pretext]{\includegraphics[scale = 0.55]{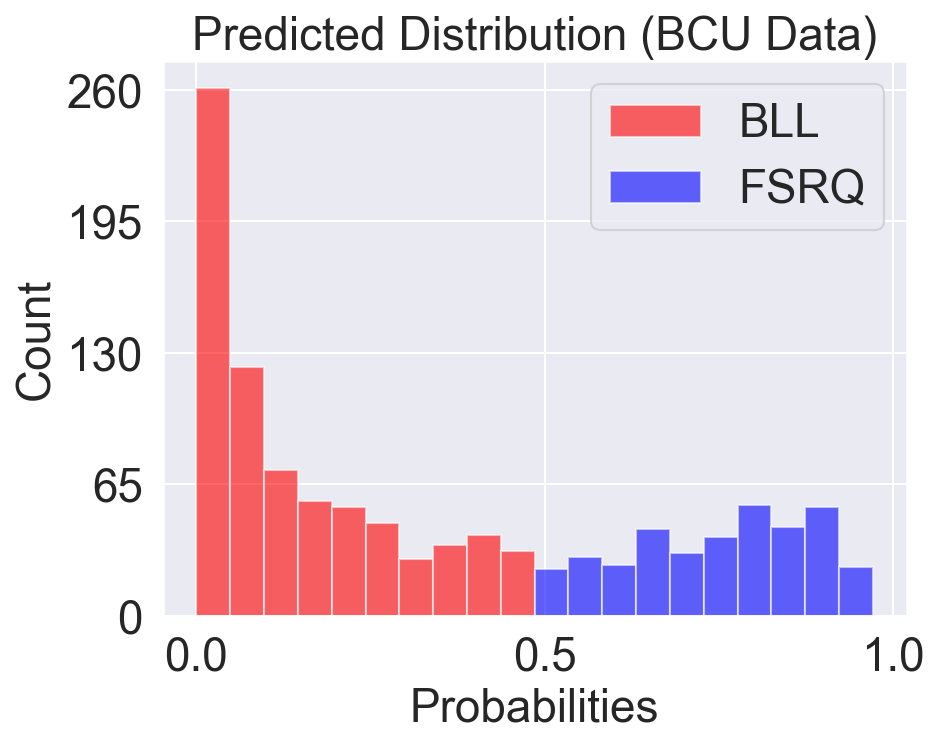}\label{fig:predicted_data_model4}} \quad
    \subfloat[SSL Contrastive Classification]{\includegraphics[scale = 0.55]{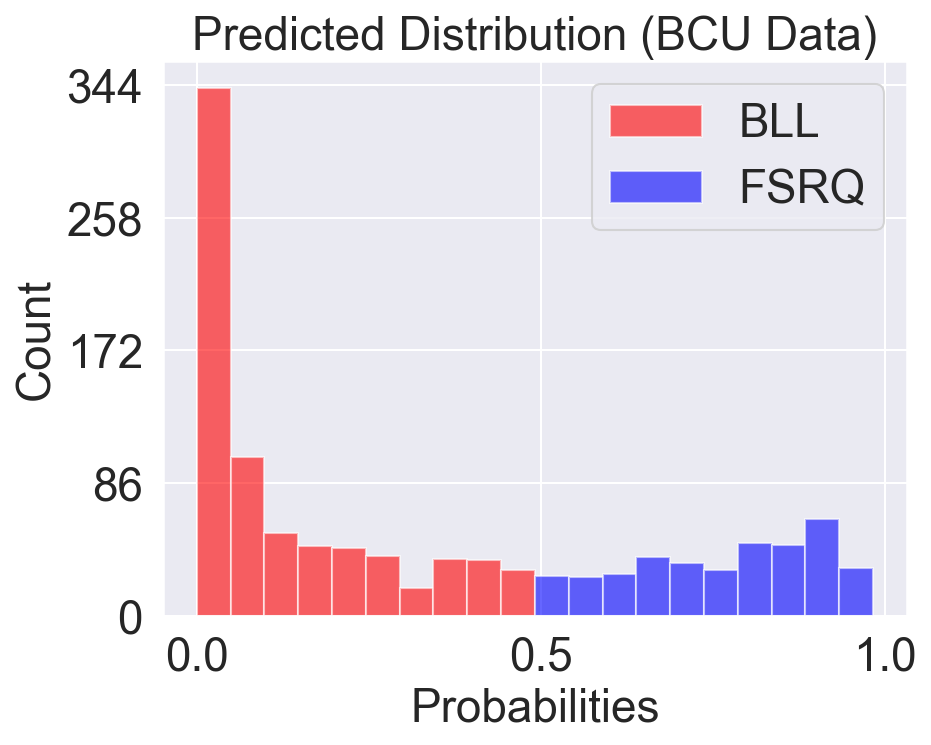}\label{fig:predicted_data_model5}}
    \caption{Distribution of Predicted output on BCU sources. Here "0" denotes the BL Lac and "1" denotes FSRQ. Again, the closer the value is to these extremes, the more confident the model is in its prediction.}
    \label{fig:Predicted_Data}
\end{figure*}

\begin{table*}
\caption{Performance Summary: A comparison table between class wise Precision, Recall, F1 score, Accuracy and AUC for all the models employed in this study}            
\label{tab:perf_summary} 
\hskip-1.7cm
\footnotesize
\begin{tabular}{ccccccccc}
\hline
Model & Accuracy & \multicolumn{3}{c}{BLLac} & \multicolumn{3}{c}{FSRQ} & AUC\\ \hline
 & & Precision & Recall & F1-score & Precision & Recall & F1-score & \\
\hline
Bias Initialisation with Soft Voting & 0.930 & 0.938 & 0.965 & 0.951 & 0.909 & 0.847 & 0.877 & 0.949 \\
Supervised Greedy Pretraining & 0.915 & 0.963 & 0.915 & 0.939 & 0.818 & 0.915 & 0.864 & 0.956 \\
Unsupervised Greedy Pretraining & 0.886 & 0.941 & 0.894  & 0.917 & 0.773 & 0.864 & 0.816 & 0.956 \\
SSL Autoencoder Pretext & 0.891 & 0.941 & 0.901 & 0.921 & 0.785 & 0.864 & 0.823 & 0.949 \\
SSL Contrastive Classification  & 0.886 & 0.941 & 0.894 & 0.917 & 0.773 & 0.864 & 0.816 & 0.951\\
\hline
\end{tabular}
\end{table*}

\begin{table*}
\caption{Performance Summary: A comparison table between macro and weighted averages for all the models employed in this study}            
\label{tab:averages} 
\hskip-1.7cm
\footnotesize
\begin{tabular}{ccccccc}
\hline
Model & \multicolumn{3}{c}{Macro Average} & \multicolumn{3}{c}{Weighted Average}\\ \hline
 &  Precision & Recall & F1-score & Precision & Recall & F1-score  \\
\hline
Bias Initialisation with Soft Voting & 0.924 & 0.906 & 0.914 & 0.930 & 0.930 & 0.930  \\
Supervised Greedy Pretraining & 0.891 & 0.915 & 0.901 & 0.920 & 0.915 & 0.917  \\
Unsupervised Greedy Pretraining & 0.857 & 0.879 & 0.866  & 0.891 & 0.886 & 0.887  \\
SSL Autoencoder Pretext & 0.863 & 0.883 & 0.872 & 0.895 & 0.891 & 0.892 \\
SSL Contrastive Classification  & 0.857 & 0.879 & 0.866 & 0.891 & 0.886 & 0.887 \\
\hline
\end{tabular}
\end{table*}

\begin{table*}
\caption{Number of BLLs and FSRQs predicted for the $1115$ BCUs by different models.}
\centering
\begin{tabular}{ccc}
\hline
Model & Number of BLLs Predicted & Number of FSRQs Predicted \\
\hline
Bias Initialisation with Soft Voting & 890 & 295 \\
Supervised Greedy Pretraining & 691 & 424 \\
Unsupervised Greedy Pretraining & 691 & 424 \\
SSL Autoencoder Pretext & 754 & 361 \\
SSL Contrastive Classification & 757 & 358 \\
\hline
\end{tabular}

\label{table:bcu_prediction_stats}
\end{table*}

The second major approach in this study is self-supervised learning, where we employ pretext tasks such as autoencoders \cite{chen2023context} and contrastive classification \citep{henaff2020data, oord2018representation, liu2023devil}. Though autoencoder is an unsupervised algorithm, it has been widely used in a self-supervised regime wherein the autoencoder is used to train the encoder to have a meaningful representation of the data. This is known as the pretext task which isn't the primary task the model is aimed at, but helps in generating a better understanding of the data for the downstream tasks. Here, the encoder's learned representations are fed as an input to a classifier for the main downstream task of classification. For our case, the encoder+decoder architecture consists of $7$ layers with input and output layers having $7$ neurons respectively. Next, the remaining layers of the encoder consist of $128$, $64$, and $32$ neurons respectively. On the other hand, the decoder has a total of $2$ layers prior to output with $64$ and $128$ neurons respectively. As this acts as a regression-based task wherein we need to reconstruct the input the loss function used here is the "mean squared error". Further, the encoded representation having $32$ neurons is directly connected to the output layer making use of "softmax" as an activation function. 

The final model that we discuss in this study is "contrastive classification". Broadly this method leverages the concept of similarity and dissimilarity between data samples. It aims to bring similar samples closer in the feature space while pushing dissimilar samples apart. For this particular case of classification, we create positive and negative pairs of FSRQs and BL Lacs, where a pair of two FSRQs or two BL Lacs is considered positive pair and a pair of two dissimilar objects i.e one FSRQ and one BL Lac is considered negative pair . The model then tries to learn how to distinguish between two types of pairs as the pretext task and while doing so it learns about the data which in turn can be used in downstream task. To do so, we create positive and negative pairs and then move ahead with creation of model, for that we define two input layers having the number of neurons equal to the feature size. Both input data samples are then passed through a shared dense layer with $64$ units and ReLU activation. Further, the embeddings of the two samples are concatenated and a dense layer with a sigmoid activation is used to predict if the pair is positive is negative. As we now have the features extracted by the constrastive learning model, we then feed it to our standard classifier having connected the extracted features directly to the output layer.

In summary, we propose $5$ different models - $3$ involving ANNs and $2$ which use self-supervised learning. The final class prediction from each model is obtained by applying a threshold of $0.5$ on the predicted probability, where all values above $0.5$ are classified as FSRQs(1) and those below are classified as BLLs(0).

\subsection{Training and Validation}
The proposed algorithms are implemented using TensorFlow \footnote{\url{https://www.tensorflow.org/}}. To ensure the reproducibility of the achieved results, we fix a random seed, hence irrespective of the number of times the algorithm is implemented, the trained weights will remain the same. Next, to avoid some over complication, we consider all the data points in a batch while training as opposed to the standard mini-batch approach to calculate the loss and accordingly optimize the model. One of the major reasons behind having done this is the dataset being heavily imbalanced, considering a smaller batch size may result in all the samples being of one single class and hence distorting the learning process. Though this can be avoided by doing a batch normalization or by smartly dealing with the sampling process, we prefer to go with a simple method resulting in minimization of computational requirements. Next, to optimize the algorithm we make use of "Adam" \cite{bae2019does,mehta2019cnn} as an optimizer which is one of the majorly used optimizers in the ML community due to obvious reasons. Note that, we do not discuss the reasons in this study as there's a large literature available on the optimizer commenting about it's features like incorporating momentum along with being a variant of Adagrad, which ultimately helps in quicker convergence. 

One of the core advantages of our algorithm is the number of parameters which is significantly less when compared to other existing models making our model faster and easier to deploy without compensating for the results. To account for rough estimate on the speed of our algorithm relative to the other algorithms, we calculate the number of FLOPs. FLOPs are generally calculated based on the number of parameters in a Neural Network model. Having fewer FLOPs is particularly advantageous for accelerating the predictions, making it easy and efficient to use in bulk. The number of FLOPs done while predicting the results is $672$, which is significantly lower than \cite{2023ApJ...946..109A} $(5056)$. As a result, the best performing model is further deployed on Streamlit \footnote{https://streamlit.io/} and Amazon AWS \footnote{https://aws.amazon.com/} so that a user can enter the values for all the features and get a prediction along with the prediction probability for the corresponding input.

Next, while training, all the models are trained for a maximum of $1500$ epochs with a stopping condition on validation accuracy, ensuring that the model will stop its training if there's no increase in validation accuracy for $300$ epochs. This introduction of stopping criteria, along with the implemented dropout of $0.5$ helps the algorithm avoid overfitting and also avoid unnecessary computations which wouldn't improve the results.

\section{Results and Discussion}
The proposed method is evaluated on the test sample which has never been seen by the algorithms during its training. One of the major reasons to have a completely independent test data is to avoid any biases that would occur if the model had updated its weights on the same data. The test data consists of $142$ BL-Lacs and $59$ FSRQs which approximately comprises of $10\%$ of the labelled data. The test data is randomly chosen from the labelled data such that it contains all the possible input distributions in a generalised fashion. 

For each model, we present a corresponding confusion matrix shown in Figure \ref{fig:confusion_matrix}. As seen in the figure, the bias initialisation with soft voting corresponds to lowest misclassification rate. Though, the number of FSRQs predicted correctly by the algorithm is the lowest compared to other models, the numbers just differs by a few samples, and hence this does not prove to be a barrier. Next, to have an appropriate metric for comparison, with the help of confusion matrix, we calculate various parameters such as Precision, Recall, F1 Score and Accuracy as shown in Table \ref{tab:perf_summary}. As we are doing a binary classification, we also calculate the Area Under the curve (AUC) score corresponding to every model as seen in the last column of Table \ref{tab:perf_summary}. To calculate AUC score, we first plot the Receiver Operating Characteristic (ROC) curve which is a plot between the "True Positive Rate" and the "False Positive Rate" as seen in Figure \ref{fig:ROC}. Next, we calculate the area under curve to identify the AUC value. Ideally, the more close the value is to $1$, the better the score is. Though from AUC score it seems that greedy pretraining is an optimal technique to proceed with, the relative difference in the score is minor compare to the missclassification error. Additionally, in the context of imbalanced data classification, F1 score is a much better metric than AUC. To validate our claim, we further calculate the macro and weighted averages of all the metrics as shown in Table \ref{tab:averages}. The major difference between macro and weighted average is that macro average is the arithmetic mean of individual scores, while weighted average includes the individual sample sizes. Macro average is calculated using the unweighted mean, which treats all classes equally regardless of their support values. This can penalize the model if the performance in minority classes is poor. On the other hand weighted average takes into account the number of true instances in each class to cope with class imbalance. As seen from the table, both the macro and weighted average F1 scores is better for the bias initialiser with soft voting. Thus considering its lowest misssclassification rate and highest weighted and macro F1 scores, the "Bias Initialisation with Soft Voting" model turns out to be the best one.

Next, we also plot the histogram corresponding to the output value given by the model on the test data. This helps in identifying if the models are overconfident in their predictions by simply comparing it's histogram with the misclassification rate. As in our case BL Lacs are denoted by $0$ and FSRQs are denoted by $1$, the magnitude of the bars close to $0$ relates to the algorithm being confident about the target being BL Lac, similar to the case with FSRQs wherein we focus on magnitudes of the bars close to $1$. As seen in Figure \ref{fig:predicted_test_model1}, compared to its counterparts, model does not gives a confident prediction for a large number of samples. It's highly probable that most of the FSRQs misclassified as BL Lacs will have a prediction value around $0.5$, and, the same is expected to be the case for BL Lacs misclassified as FSRQs. On the other hand for all the other algorithms particularly for the self-supervised ones we see a very high magnitude bar around the value $0$, indicating that model is confident about most of the samples being BL Lacs, and the same has been observed from the confusion matrix. However, an important point to note here is that, these algorithms tend to classify a lot of BLLs as FSRQs and hence being overconfident about wrong predictions. Thus, even though the flatter distribution in \ref{fig:predicted_test_model1} might seem counter-intuitive to well separated categories, it reflects the fact that this model is not disproportionately biased to either class. The comparatively higher peaks around $0$ for models \ref{fig:predicted_test_model3}, \ref{fig:predicted_test_model4} and \ref{fig:predicted_test_model5} indicate that they are biased towards BLLs since, they identify a considerably lesser number of BLLs correctly than \ref{fig:predicted_test_model1}.

In addition to its optimal performance, one of the major advantages of our "Bias Initialization with Soft Voting Neural Network" is the model's parameter count, as discussed in the previous subsection. Recently, numerous studies have explored similar approaches \citep{2023MNRAS.525.1731C,2023ApJ...946..109A,2023MNRAS.519.3000S}, employing a large number of features and are thus, limited to a small number of samples, due to a high percentage of missing values in the catalogue. On the other hand if the algorithms estimate these missing parameters e.g., redshift before classification, then their results will be heavily biased based on the imputation technique used. In contrast, Agarwal's optimal model shares the same number of features as ours. However, the considered model in the work requires approximately seven times more FLOPs (Floating Point Operations) compared to the presented model. This computational efficiency enhances the user experience when using our tool without compromising on results. You can find a link to the web app in the Data Availability section.

\begin{figure*}
    \centering
    \subfloat[Distribution of Photon Indices in the Predicted Sample]{\includegraphics[scale = 0.5]{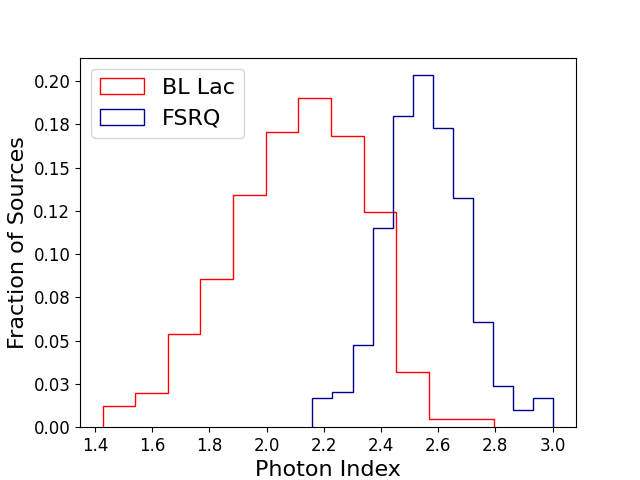}\label{fig:photon_index}} \quad
    \subfloat[Photon Index Distribution across Logarithmic Pivot Energies in the predicted Sample]{\includegraphics[scale = 0.5]{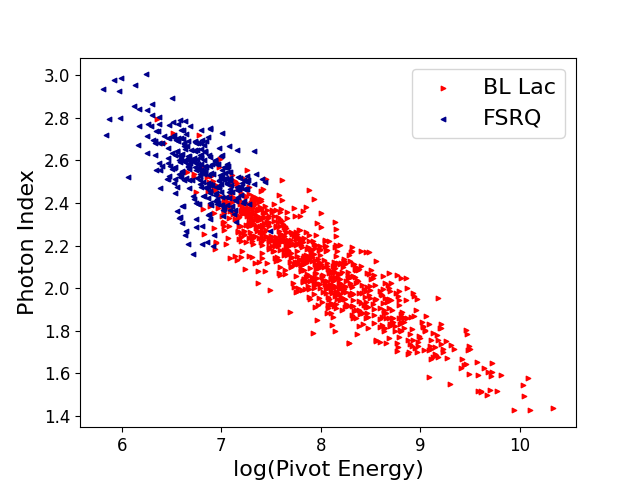}\label{fig:scatter}}\quad
        \caption{The dichotomy between BL Lac and FSRQs is evident in their distribution within the predicted sample. }
    \label{fig:characteristic}
\end{figure*}

We make use of each model to make predictions for BCUs \footnote{https://github.com/abhimanyu911/bcu-classification/tree/main/bcu\_predictions}. The histograms can be seen in Figure \ref{fig:Predicted_Data}. Out of $1115$ BCU samples, the optimal algorithm ("Bias Initialization with Soft Voting") of ours predicts that there are $820$ BL Lacs and $295$ FSRQs, the list of which is made available in our GitHub repository. This number can be considered approximate due to the uncertainty and errors associated with every ML model. One of the other supporting evidence for these numbers is the redshift associated with the BCU samples. In our previous paper, we observed that a significant number of BCUs in the catalog has a lower redshift (Please refer to Figure 4 and Figure 5 in \citealt{gharat2023estimation}) indicating that there would be more BL Lacs compared to FSRQs which perfectly aligns with the claim made in this study. Table \ref{table:bcu_prediction_stats} presents the predictive statistics for the unclassified BCUs according to each model. Notably, there is a separate consensus between the greedy pretraining approaches and between the self-supervised learning techniques, with each of the 5 models concurring that no less than $60$\% of these samples are likely BLLs.

We observe a significant dominance of BL Lacs in the resulting sample, making up approximately 73\%, which is consistent with the observation in the 4LAC catalogue, where the number of BL Lacs is nearly double that of FSRQs. This dominance is further supported by the challenges posed by a large number of BL Lac objects displaying weak or no emission lines, making the detection of their optical counterpart information difficult and classification a complex task. Consequently, it is reasonable to infer that a significant fraction of the BCUs in the Fermi/LAT catalog are highly likely to be BL Lac objects. Also, it is important to note that considering the binary nature of the classification, limited to BL Lac and FSRQs, and excluding other potential classes such as Seyferts, radio galaxies, and other AGN -- constituting  only a small fraction of the sample -- we anticipate that less than 3 percent of non-blazar AGN subclasses could potentially introduce contamination in the BCU sample.

To explain the apparent dichotomy between BL Lacs and FSRQs, the intrinsic difference in the  nature of the accretion disk and physical origins of $gamma$-ray emissions in BL Lacs and FSRQs may contribute to their distinct gamma-ray properties, such as the larger gamma-ray luminosity of FSRQs and the harder gamma-ray spectral characteristics of BL Lacs. For instance, BL Lacs can have a \emph{radiatively inefficient accretion flow} and be more magnetically dominated, resulting in a lower mass accretion rate compared to FSRQs \cite[e. g.][]{2009MNRAS.396L.105G,2011MNRAS.414.2674G,2019MNRAS.486.3465M}. Similarly, the absence of circumnuclear gas near the central engine, as indicated by the featureless optical spectra of BL Lacs, suggests that the origin of gamma-ray emission in the sources can predominantly be ascribed to SSC emission \citep[e. g. ][]{1997A&A...320...19M}. On the other hand, studies of the broadband SED of FSRQs suggest that gamma-ray emission in FSRQs is mainly contributed by EC \citep[see e. g.][and reference therein]{2015ApJ...807...79H}.

To ensure the model's consistency in accordance with the characteristic features of the two classes, we plotted the distribution of the photon index. We observed that the BL Lacs in the predicted sample exhibit a relatively harder photon index compared to FSRQs, as illustrated in Figure \ref{fig:photon_index}. (see also Figure 1 and 3 in \cite{2022ApJS..263...24A}. Additionally, we also plotted a graph between photon index and log of pivot energy and observed results align with the expected trend, positioning BL Lacs in the upper left of the anti-correlation trend, as seen in Figure \ref{fig:scatter} (see also Figure 2 in \cite{2019ApJ...887..134K}). The results from the study can be further verified and refined through dedicated multi-wavelength observations, involving telescopes ranging from radio to TeV. Moreover, as BL Lacs represent the most extreme class of AGN, this sample can serve as the target sample for future ground-based TeV and PeV telescopes.

\section{Conclusion}
The recent catalog from the Fermi/LAT gamma-ray telescope contains a large number of AGN sources that require decisive classification. This is because classification using gamma-ray data alone is not always possible.
In this study, we employ multiple algorithms to classify blazars of unknown class into BL Lacs and FSRQs.
The proposed models exhibit simplicity in their nature, with their distinctiveness stemming from the fine-tuning of the method through appropriate initialization and the incorporation of soft voting. Additionally, we delve into a couple of self-supervised algorithms in their vanilla form to assess their capabilities. "Bias initialization with Soft Voting" emerges as the best-performing model in our case. The algorithm's ability to make a larger number of predictions can be attributed to the minimal number of features it utilizes, which stands as a key factor. Furthermore, our study places emphasis on maintaining a minimal number of parameters while still delivering strong performance, which proves to be a pivotal feature. This enables us to deploy the model in various scenarios.

 \section*{Acknowledgements}
We thank the anonymous referee for their careful and thorough review of this paper, which helped improve the quality of the work


\section*{Data Availability}
The data utilized in this paper can be accessed by the public through the Fermi Science Support Center (FSSC) of NASA's Goddard Space Flight Center. Following the reproducibility and open source standards followed by the ML community, we make all our codes public that can be accessed through our \href{https://github.com/abhimanyu911/bcu-classification}{Github repository: (https://github.com/abhimanyu911/bcu-classification)}. Further, to have a comfortable experience with our method, the best performing model is deployed on \href{http://13.239.10.157:8501/}{AWS: http://13.239.10.157:8501/)} and \href{https://bcu-classification-ml.streamlit.app/}{Streamlit: https://bcu-classification-ml.streamlit.app/}. Note that the AWS app may get deactivated after the expiry of credits, in such cases we recommend a user to make use of Streamlit. In case of any issues, a docker image can be provided on reasonable request to sarveshgharat19@gmail.com.



\bibliographystyle{mnras}
\bibliography{example} 








\bsp	
\label{lastpage}
\end{document}